# GeoAI in resource-constrained environments

Gede Sughiarta
Wisnu foundation
Denpasar, Indonesia
gede@wisnu.or.id

Atiek Kurnianingsih
Wisnu. foundation
Denpasar, Indonesia
atiek@wisnu.or.id

Srikar Reddy Gopaladinne
Department of Computer
Science and Engineering
University at Buffalo
Buffalo, USA
srikarre@buffalo.edu

Sujay Shrivastava
Department of Computer
Science and Engineering
University at Buffalo
Buffalo, USA
sujayshr@buffalo.edu

Hemanth Kumar Reddy Gorla
Department of Computer
Science and Engineering
University at Buffalo
Buffalo, USA
hgorla@buffalo.edu

Marc Böhlen
Department of Art
Institute for Artificial Intelligence
and Data Science
University at Buffalo
Buffalo, USA
0000-0002-0463-0988

*Abstract*—This paper describes spatially aware Artificial Intelligence, GeoAI, tailored for small organizations such as NGOs in resource constrained contexts where access to large datasets, expensive compute infrastructure and AI expertise may be restricted. We furthermore consider future scenarios in which resource-intensive, large geospatial models may homogenize the representation of complex landscapes, and suggest strategies to prepare for this condition.

## I. INTRODUCTION

Spatially aware Artificial Intelligence, *GeoAI*, is changing the way the surface of Earth is observed. However, these changes and advancements are accompanied by significant costs in the procurement of data, computing resources, communication infrastructure and human expertise. Moreover, many ambitious, large GeoAI models are trained for land cover categories that are mostly of interest in the Northern Hemisphere. Given this imbalance, we ask how participants from low-resourced environments can best make use of GeoAI. We contextualize the discussion with a case study in Central Bali in which we developed multiple GeoAI approaches to assist the Wisnu foundation, a non-governmental organization (NGO), founded in 1993, in their ongoing efforts to manage community resources and to perform land mapping of small villages in Bali. With this case study as a backdrop, we reflect on the conditions and limitations of deploying complex GeoAI algorithms in resource-constrained environments and propose a combination of GeoAI pathways, of varying technical complexity, as a way to respond to obstacles preventing low-resource organizations from benefiting from GeoAI opportunities.

Land cover analysis typically relies on low-earth orbit satellites to collect geospatial datasets, which are then analyzed to detect change across time. From changes in land cover, changes in land use can be selectively inferred. These procedures are applied to detect variations in the level of urbanization, in agricultural land use and in the extent of forest cover [1], for example. Such remote sensing-based land cover analyses have been conducted at a variety of scales over the last few decades [2]. However, not all regions of the planet are equally primed to profit from these GeoAI activities. Even simple operations can be hampered by technical and institutional obstacles, as other researchers have observed [3].

As knotty as institutional limitations can be, additional barriers are present in low-resourced environments outside developed countries. Established land cover definitions, such as the European Union's *CORINE* inventory [4], are particularly focused on landscapes contained within the European Union, but do not cover territory such as tropical rainforest and desert conditions in detail. Moreover, Western land management-centric landscape analysis can lead to misreading of land use in the majority world [5]. These misinterpretations can be due to indifference to nuance, lack of literacy in reading unfamiliar landscape conditions, or biases in evaluating landscapes that are not optimized for extractive land-management operations.

In the past, land cover analysis operations were undertaken mostly by nation states, large companies or international organizations [6]. Small groups, such as NGOs, have made less use of remote sensing-based land mapping due to the costs associated with the process. The availability of publicly-sourced satellite imagery from the European Union such as the Sentinel-2 satellites [7] as well as the availability of web-centric visualization environments [8], have lowered the bar to working with satellite-sourced datasets. Moreover, the recent addition of private-sector sourced high spatial resolution imagery with unprecedented temporal resolution has opened the door to observation regimes previously reserved for military applications. While the prospect of the *daily change map* [9] is alluring, it remains wishful thinking for most parts of the world, either due to the costs of acquiring and the complexity of processing the requisite data, or because atmospheric disturbances, such as cloud cover, render some areas of the globe less accessible to visual inspection than others.

While the expansion of remote sensing products improves access to novel observation resources, they also, perhaps counter–intuitively, limit what can be performed with them. With a focus on making complex processes easier and streamlining data-intensive operations come limitations that can prevent non-experts, such as NGOs from using the resources for their own interests. The next sections of this paper explore just this problem through the lens of the Wisnu foundation. In order to better understand how the new GeoAI landscape expands and limits the foundation's mission, we will review the most important approaches to land cover analysis while considering their applicability to low-resource environments.

## II. OVERVIEW OF GEOAI

GeoAI represents the convergence of geospatial data analysis and AI, enabling advanced spatial data processing and interpretation. While GeoAI has been energized by the ongoing AI intensification, AI in Geography has been a topic for several decades [10]. More recently, the focus has been set on AI's capacity for knowledge discovery in spatial inquiries in light of deep Neural Networks. These networks are large models and super large spatial data collections, resulting in spatially explicit models with "spatial representations of the studied phenomena" [11]. The integration of spatial conditions into AI analysis pipelines has resulted in GeoAI enabled contributions to urban planning, environmental monitoring, disaster management [12], smart cities design [13], deforestation [14], flood susceptibility modelling [15], wildfire detection [16], Arctic iceberg tracking [16] and even illegal industrial-scale fishing monitoring [17].

## III. GEOAI ALGORITHMS FOR LAND COVER ANALYSIS

Land cover analysis is a form of spatial image processing and typically relies on semantic segmentation to detect land cover conditions. Semantic segmentation assigns each pixel in an image to a predefined category. This assignment is dependent on the specifics of a given algorithm, ranging from color and intensity of a pixel to more complex operations that include patterns surrounding a pixel. Land cover analysis predates the advent of GeoAI and has been, similar to other data-heavy engineering topics, re-invigorated by Neural Network-centric GeoAI procedures. The most widely used approaches to image segmentation are Random Forests (RF), Support Vector Machine (SVM), Neural Nets (NN), U-Nets and Transformer Networks (TN). The next paragraphs briefly describe each of these algorithms in order to compare their applicability to our particular use case.

RF [18] methodically partitions a dataset into subsets and assigns them to a collection of decision trees. Each tree assesses data using various criteria. After each tree reaches its conclusion, RF combines their outputs through a voting process to determine the most prevalent class, producing a final prediction.

SVM [19] is a powerful algorithm for identifying clusters in data by strategically placing separation planes between data points, effectively dividing them into distinct groups. It seeks an optimal hyperplane that maximizes the margin between different data points in a dataset, making it robust even in high-dimensional spaces.

NN [20] consist of interconnected nodes that store and transmit data. Each node's signal strength, controlled by assigned weights, influences subsequent layers. Typically, NNs include an input layer receiving external data and an output layer providing the network's response. Between these, there can be multiple - often hundreds – of layers. During operation, data samples are processed, and an objective function calculates the error between current and desired outputs. Through backpropagation, the network adjusts its many internal parameters to minimize this error until a stopping criterion is met. For tasks such as image analysis, NNs preprocess data using techniques such as convolution operations. Variations of NN systems have been developed to address specific deficiencies of NN systems for image-based GeoAI operations. Residual Networks (ResNets) are NNs that tackle training issues in deep models by introducing shortcuts that combine input features directly with those from adjacent layers.

U-Net [21] U-Nets are structured with two interconnected computational paths: one emphasizing context and a symmetric counterpart focusing on localization. Unlike traditional convolutional networks that typically reduce dimensionality, U-Net integrates upsampling operations to enhance output resolution. It combines high-resolution features from the contracting path with the upsampled output, promoting more accurate predictions.

TN represent a recent addition to the suite of algorithms applicable to land cover analysis. Initially designed for natural language processing, TNs excel in handling sequence-to-sequence tasks over long-range dependencies using the attention mechanism [22]. Vision transformers extend this approach to imagery by dividing images into patches, transforming them into a linear sequence of embeddings for standard transformer encoders. This adaptation allows vision transformers to effectively encode spatial information in images. For satellite imagery, specialized vision encoders can further incorporate temporal and spectral data to enhance analysis capabilities.

NN and TN algorithms build the basis for state of the art Geography-aware supervised learning [23] that reconfigure learning frameworks for the unique properties of temporal geospatial datasets and can detect landscape conditions that earlier methods could not identify.

## IV. A CASE STUDY IN THE ALAS MERTAJATI OF CENTRAL BALI

Before we discuss the merits and problems of building end-to-end GeoAI pathways for land cover analysis and land use interpretation in resource constrained contexts, we will describe how several of the algorithms outlined above performed in a field study in the Alas Mertajati of Central Bali, depicted in Fig. 1. The Wisnu foundation has been actively supporting indigenous groups in documenting land use and environmental conditions and has, to date, conducted manual mapping operations in 28 traditional villages across the island of Bali. While skilled in cartography, landscape surveying, and interview techniques, Wisnu has not yet utilized GeoAI tools.

For Wisnu, GeoAI represents a novel approach to assess the landscape's carrying capacity for sustaining traditional Balinese mixed agrarian practices and provides an efficient means to bolster community resilience amid economic growth pressures exacerbated by tourism development, which annually converts approximately 1000 hectares of land for tourism-related purposes. Because satellite-based observation and GeoAI analytics can be streamlined once established, they can assist the Wisnu foundation in more rapidly responding to all forms of change, including climate crisis-induced change. These local capacities and initiatives are all the more significant as government-led efforts have been slow to respond to the opportunities presented by GeoAI.

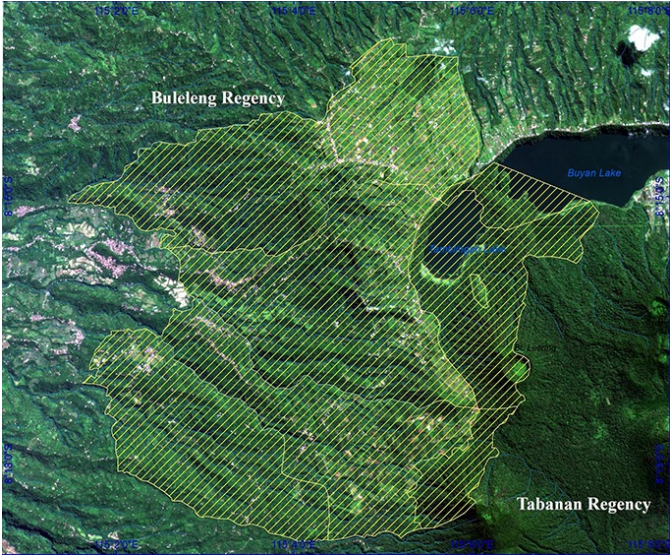

Fig 1. The Alas Mertajati surrounded by the Buleleng and Tabanan Regencies in Central Bali. Background image: Planet Labs, Superdove, May 2022.

## V. Dataset collection and GeoAI data wrangling

Between 2022 and 2023 we collaborated with the Wisnu foundation to collect extensive field data in the lands of the Indigenous People of Dalem Tamblingan (Masyarakat Adat Dalem Tamblingan) and the Alas Mertajati in Central Bali [24]. This study was conducted amidst a persistent conflict between the Tamblingan community and the Balinese government regarding forest and water resource management. A key element of the land care debate centers on sustainable land use practices. The Indigenous People of Dalem Tamblingan have long practiced agroforestry, a small-scale sustainable and largely manual-labor dependent form of land use that is adaptable and much more robust against climate fluctuations than monoculture farming. Agroforestry in the tropics combines woody perennials such as trees, shrubs, palms, and bamboos with agricultural crops and animals in unique temporal and spatial arrangements. Active agroforestry plots are indicators of traditional sustainable agricultural practices persisting into the present, and documenting these areas delivers evidence to back the case of sustainable land practices vis-à-vis local government as well as international agencies. However, the detection of agroforestry is a technically challenging GeoAI operation that relies on extensive field samples and high-resolution satellite imagery.

As opposed to textual data that can easily be pre-processed and ingested into Neural Net compatible input, remote sensing data collected from satellites require more elaborate preparation steps. This preparation often entails cumbersome manual labelling that transform identified landscape features into labeled polygon areas as illustrated in Fig. 2 below. Identifying distinct land cover features is non-trivial and requires expertise in parsing satellite imagery. In our study, we collaborated with an expert in remote sensing of tropical land conditions from the Indonesia National Research and Innovation Agency. Additionally, we consulted with local experts from the Wisnu foundation and confirmed many of these polygon data selections in the field.

These labeled polygons in turn deliver the actual input data to an image segmentation algorithm. Mean and standard deviation values are collected from randomly selected pixels contained within these polygons across all bands of a given satellite image. Some GeoAI procedures, such as the U-Net, require converting polygons into raster images, where each pixel is receives a unique label corresponding to a specific land cover category (mixed forest, settlement, agroforestry, etc.). This process creates a ground truth image for model training. The supplementary materials listed below contain further details on the U-Net specific preprocessing steps.

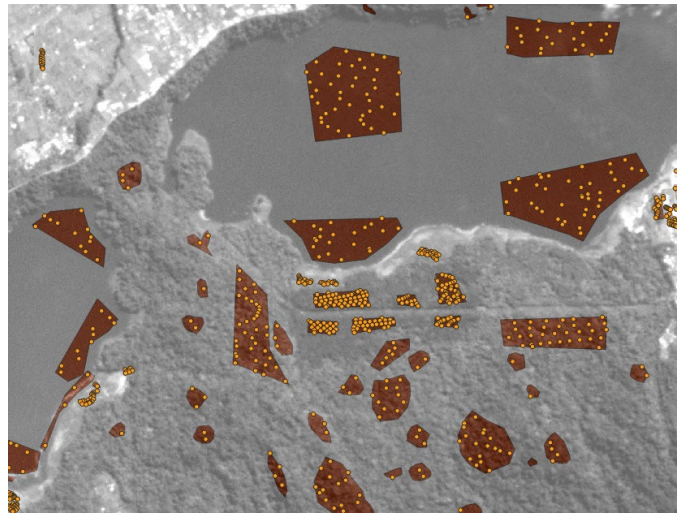

Fig 2. Land cover categories (darker areas) manually identified by domain experts and marked with polygons on the reference satellite image of May 2022. Sample pixels randomly selected from within those polygons (lighter dots) deliver statistical descriptors of the respective land cover category.

## VI. Parameter tuning and data analysis

We worked with the well-known ORFEO machine vision library [25] for the smaller GeoAI models RF and SVM and built a bespoke U-Net based on the TensorFlow machine learning library [26], utilizing the Python bindings of both systems. Through trial and error we adjusted the internal parameters of our candidate algorithms to optimize the identification of agroforestry sites while maintaining good results across all other pertinent land cover categories as described in previous reports [27]. By and large, the default parameters were close to our selection set. The SVM approach relied on a linear kernel and a C parameter of 1.0. The RF approach operated with 100 trees and 10 features only. The iterative SVM and RF algorithms terminated at a tolerance epsilon of 0.01.

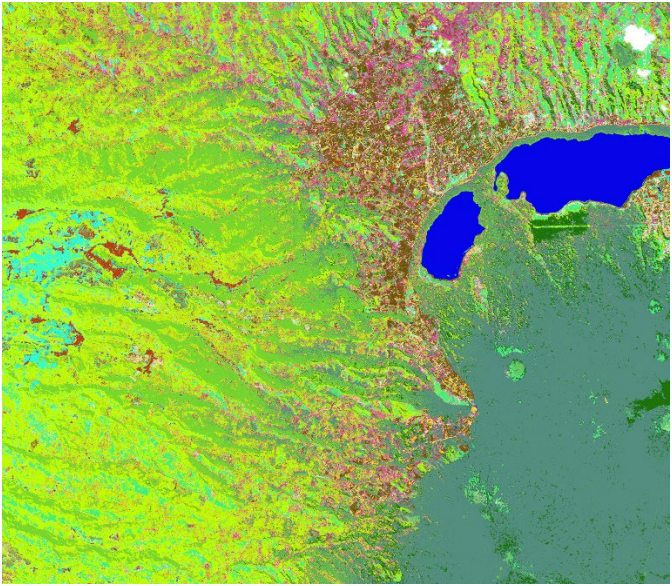

Fig. 3. SVM result. The SVM based segmentation is produced with the Orfeo machine vision library. Agroforestry is mapped in pink.

A large satellite image cannot be fed directly into the U-Net model. Instead, the 8-band satellite image is first divided into smaller 16x16 pixel patches across all bands. The U-Net model itself consists of a contraction path that extracts and down-samples image features through 5 blocks of successive convolutional layers with dropout, followed by a symmetric expansive path that upsamples and combines these features to reconstruct the image. The final layer produces a pixel-wise classification map of the same size, where each pixel is labelled according to a predefined categorical schema (e.g., mixed forest, settlement, agroforestry, etc.). These classified patches are then stitched back together to reconstruct the full image, as shown in Figure 4. The model was trained for 1000 epochs, achieving the F1 score displayed in Table 1.

Because of the variations inherent in agroforestry plots, the reliable detection of agroforestry signatures in satellite imagery is a challenging technical problem. However, RF, SVM and U-Net approaches were all able to detect, to different degrees, agroforestry in 8-band, 3-meter spatial resolution datasets provided by Planet Labs [28]. Both RF and SVM were able to perform reasonably well with comparatively small input sample sizes. This condition was particularly helpful for the analysis rice paddies and mixed garden plots that can appear in various forms (conditions 1,2,3 in Table 1) across planting and harvest events in tropical climates. However, we found that overall, the SVM algorithm was most responsive to the size limitations of our dataset while producing useful results across most of our categories. Specifically, the LibSVM version of SVM, an efficient and open source version [29] of the SVM algorithm, proved effective. LibSVM delivered good results across all 20 land cover categories and the adequate F1-scores for the complex agroforestry class. Finally, SVM was more responsive to updates with additional field data than our other candidate algorithms.

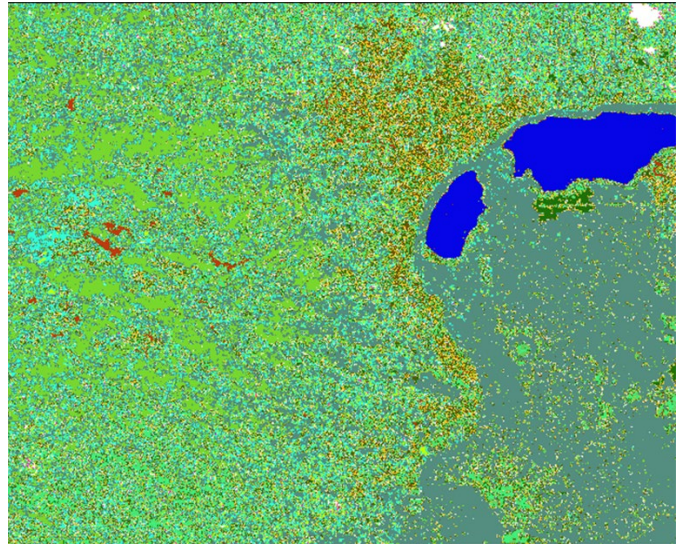

Fig. 4. U-Net result. The U-Net approach is based on Keras and TensorFlow machine learning libraries. Agroforestry is mapped in pink.

The state of the art U-Net-based approach is technically the most involved of our analysis options and demonstrated both advantages as well as some disadvantages in our use case. U-Net is one of the most effective algorithms for land cover analysis and scales much better than the traditional algorithms RF and SVM. This higher accuracy likely reflects the U-Net's superior representational capabilities, which can better model nuances and minority classes, including agroforestry plots.

While U-Net's F1 accuracy for agroforestry and related land cover categories exceeds that of RF and SVM, the result is likely an overestimate of the actual extent of agroforestry. We believe this to be the case because the U-Net architecture ingested patch sizes exceed the size of some of the smaller agroforestry plots observed in the field. The choice of the input patches was a function of the dimensions of the U-Net architecture, which in turn were selected to minimize loss during training across all land cover categories.

Figures 3 and 4 show the results from SVM and U-Net image segmentation. Further details regarding the significance of agroforestry, the process of documenting its presence in the lands of the Indigenous People of Dalem Tamblingan are described elsewhere [27].

TABLE 1

Comparison of quantitative performances expressed as F1 scores of RF, SVM and U-Net across land cover categories with spectral signatures similar to the target land cover feature agroforestry. Best results across all test runs.

| Landcover type | F1 RF | F1 SVM | FI U-Net |
|---|---|---|---|
| Clove plantation | 0.798 | 0.930 | 0.990 |
| Mixed forest | 0.654 | 0.918 | 0.895 |
| Rice paddy – condition 1 | 0.605 | 0.810 | 0.952 |
| Rice paddy – condition 2 | 0.467 | 0.771 | 0.982 |
| Rice paddy – condition 3 | 0.808 | 0.808 | 0.957 |
| Mixed garden – condition 1 | 0.707 | 0.806 | 0.993 |
| Mixed garden – condition 2 | 0.907 | 0.935 | 0.967 |
| **Agroforestry** | **0.224** | **0.603** | **0.820** |

*A. AI pathways and resource constrained operations*

While our project considered the performance of specific algorithms for land cover analysis, our actual interest targeted the overall utility of the approaches under the resource-constrained context we operated under, and the role various resource constraints play in the determination of an optimal solution. The utility of a given land cover detection algorithm would typically be evaluated based on its quantitative performance metrics alone. However, resource constrained environments call for additional considerations. Under resource constraints, factors such as data availability, expertise, compute requirements, energy use and cost weigh more heavily than in resource-rich contexts. Access to cloud compute environments is particularly effective where connectivity is robust, and bandwidth limitations are negligible. In many parts of the majority world, including Indonesia and the island of Bali, connectivity remains compromised. Additionally, while some remote sensing satellite datasets are publicly available, others are not. Sentinel-2 datasets from the European Space Agency are readily available, but the spatial and temporal resolutions are inferior compared to current private sector satellite products. The detection of agroforestry, for example, relies on the spectral and spatial information richness contained across the 8 bands of the Planet Labs Superdove satellite constellation. Small agroforestry plots typical of the highlands in Bali cannot be readily detected with Sentinel data assets.

To be clear, the costs associated with adopting GeoAI can be substantial. A significant factor in the adaptation is the ability to tune a given GeoAI approach to the specific needs and interests of an organization. Systems that facilitate computation but do not facilitate modification will be less responsive to this important requirement. Finally, AI expertise forms a crucial part of the equation. In cases where the requisite expertise is scarce, simpler solutions that align with an organization's existing operational infrastructure may be more effective than cutting-edge technologies.

Despite these structural impediments, NGOs such as Wisnu may profit from GeoAI in unexpected ways. Wisnu possesses a valuable yet often overlooked asset: intimate connections with the communities it serves and a deep understanding of the lands they inhabit. This local knowledge uniquely positions Wisnu and other NGOs to gather comprehensive and high-quality field data. The land cover algorithms evaluated in this survey are supervised machine learning algorithms and require significant amounts of high-quality data. NGOs such as Wisnu can have considerable leverage over the outcome of supervised machine learning operations precisely because of their access to local expertise and intimate knowledge of ground conditions, cultures, and contexts. At the same time, collecting and curating such data is a time-intensive and expensive process. Moreover, the expertise required to prepare datasets for ingestion into machine learning pathways is typically not a skill an NGO holds at the ready. All of these divergent factors have to be considered when evaluating the best approach to making GeoAI applicable to resource restricted environments. In response to these conflicting GeoAI opportunities, we experimented with several different combinations and workflows, all utilizing the same curated datasets. These different GeoAI pathways allowed us to consider basic tradeoffs across algorithms and workflows that can generalize to other contexts, despite the fact that the field of GeoAI remains very much in flux. Here are the four different GeoAI pathways we identified and made use of.

*1) Local agency, small datasets and low computational complexity*

This pathway includes the use of a small model, created for example with the RF algorithm available through the QGIS [30] framework, performed by local Wisnu foundation members, in consultation with support from the externally operating research team. This approach generates a good first approximation of core land cover categories and can be easily iterated on with additional field data. The RF algorithm available within QGIS can execute on low-end hardware, such as older laptops and desktop computers. Some disadvantages of this first pathway include the facts that the RF algorithm itself cannot be modified outside select hyperparameters, and that the QGIS environment often requires manual adjustments across the various add-on components, including controlling for library version conflicts as not all subsystems update synchronously. Additionally, this pathway does not readily scale to larger datasets and more complex models.

*2) Local agency, small datasets and higher computational complexity*

While at least three of our evaluated algorithms offer useful results, the best combination of accuracy and ease of use occurred in the combination of SVM on cloud compute infrastructure. We created a bespoke cloud-enabled environment, Cocktail (see supplementary materials), that made use of the ORFEO machine vision library to implement a LIBSVM version of SVM. However, the Cocktail environment requires command-line interaction, a process that non-engineers often find cumbersome and restrictive. Nonetheless, the environment proved invaluable for the remote research team members in several ways, including the ability to easily share parameter settings across training operations, as well as the opportunity to facilitate the comparison of results. Because Cocktail sends its output directly to a connected low-cost server, results can easily be shared, including with partners not directly involved in the code-centric analysis process. Making GeoAI as well as the artifacts it produces easily understandable and shareable is particularly important where the integration of viewpoints and positions across multiple stakeholder groups is desirable.

*3) Computational flexibility, scalable models, larger datasets*

Our third approach focused on the previously described bespoke U-Net model that can operate in a variety of contexts, from a modest CPU system to hardware-accelerated GPU configurations. As mentioned above, a U-Net model built with the same dataset used for SVM yielded useful analytic results for the agroforestry land cover category. Also, this pathway supports scaling if computational resources are available. Our U-Net scripts respond to this opportunity in part as notebook versions for stepwise experimentation and visual evaluation, optimized for cloud-centric use. The bespoke U-Net approach offers more flexibility, as we can create larger or smaller models with more or less internal model parameters to respond to the underlying data conditions. Moreover, the bespoke approach allows us to modify the model optimization process, and thus to tune for a specific land cover category, for example. However, the scaling advantage of U-Net cuts two ways. The opportunity to incorporate more data and larger models increases the cost of generating results. Where datasets are proprietary, as is the case with Planet Labs imagery, the advantage of larger and more representative data intensive models may become prohibitive. For that reason, this pathway is best considered a long term and expensive investment that supports a cutting-edge deep learning GeoAI approach while retaining some of the technology development in the hands of local stakeholders.

*4) Canonical datasets and pre-trained (foundation) models*

This last category is significant as it combines large general-purpose datasets with very large NN models. As datasets themselves are significant components of the combined effort, it is important to point out some of the unique features of these large data collections, and to describe their impact on the large models they inform.

A case in point is the Functional Map of the World (FMW) [31], a dataset with 63 distinct categories that aims to support the creation of machine learning models capable of predicting functional land use from temporal sequences of satellite images and associated metadata. SatlasPretrain [32] has a similar goal and is more ambitious yet. It combines Sentinel-2 and NAIP imagery [33] with over 300 distinct labels and 137 classes represented in 64 million images. Both of these datasets are important contributions to GeoAI, and yet they carry significant baggage. Even a cursory view of the categories included in the FMW illustrates a specifically goal-oriented mapping of landscape. For instance, FMW includes three unique features of airports and seven unique descriptors of energy processing facilities. The organization scheme of FMW is perhaps similar to the sometimes tone-deaf logic of ImageNet [34] that long operated as a canonical source for training image classifiers. Consequences from reductionist representation of socially salient categories in ImageNet, including representational biases, emerged after systems trained on ImageNet were deployed in the field, such as in interview assessment tools [35].

New large NN models require enormous amounts of training data and are designed to take advantage of the new and large geospatial data collections. For example, both SatMAE [36] and SatlasPretrain make large pre-trained models available either in generic form or as base models that can be fine-tuned to specific applications. However, SatMAE and SatlasPretrain differ in their model configurations. SatMAE is designed with Masked Auto Encoders, while SatlasPretrain operates on ResNet architectures. In both cases, the resultant models are large. For example, the SatMAE model (VIT large) is 400 MB, the SatlasPretrain models approximately 200 MB, while the bespoke U-Net amounts to only 8 MB. Moreover, SatMAE requires fine-tuning on 8 NVIDIA V100 GPUs, and the SatlasPretrain models can practically only be processed on cloud-centric compute systems which remain mostly out of reach for small organizations in the majority world. Perhaps the most important and largely invisible impediment to making use of this category of large pre-trained models are the built-in biases and preferences regarding landscape categorizations as illustrated in Figures 5 and 6.

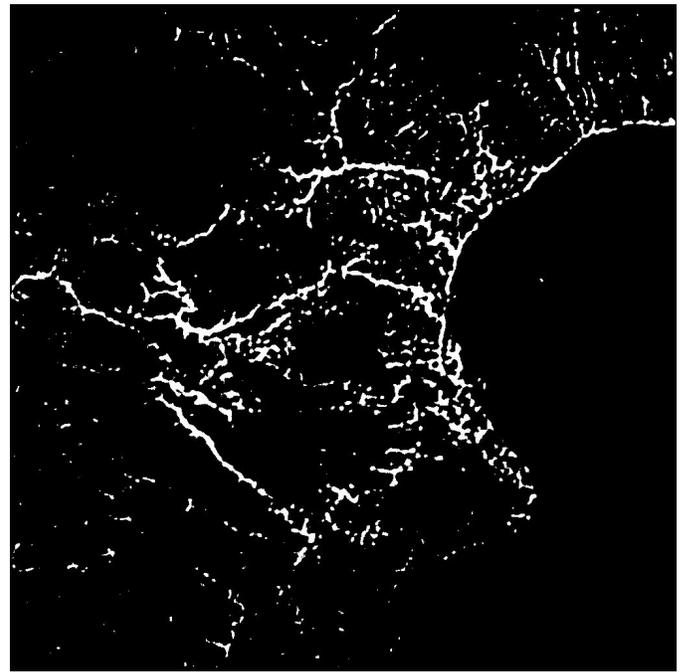

Fig. 5. SATLAS experiments on Sentinel-2 data of the study area with pretrained Resnet152 model. White areas indicate roadways.

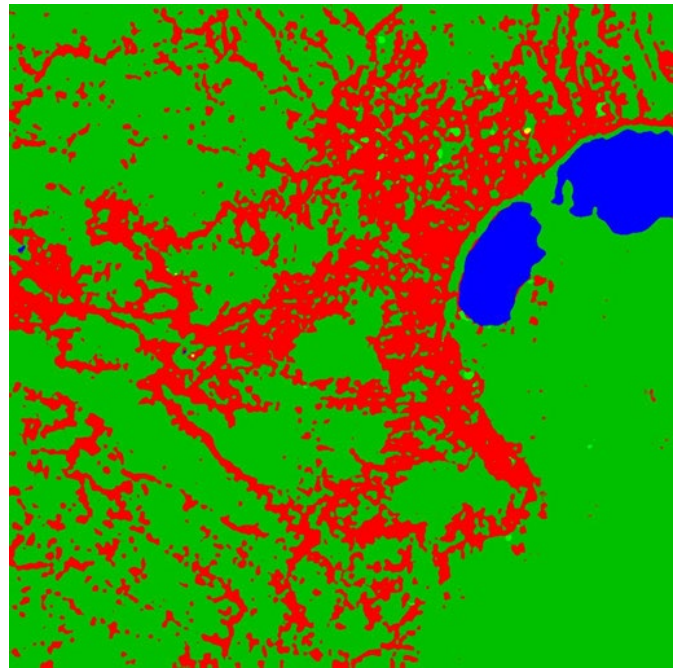

Fig. 6. SATLAS experiments on Sentinel-2 data of the study area with pretrained Resnet152 model. Water (blue), developed (red), tree, shrub, grass (green). Agroforestry and rice paddies remain invisible.

SatMAE bases its definition of land cover categories on the Functional Map of the World classes that emphasizes, as alluded to above, industrial landscape features and does not include small scale agricultural land cover conditions. SatlasPretrain, for its part, is based on a more differentiated collection of land cover conditions and crop types that are of interest to inhabitants of the majority world, including coffee, sugarcane, rice, and cassava. In order to evaluate how well the generic Satlas models might perform in our particular context, we applied the Satlas pretrained model ResNet152 to a stack of eight Sentinel-2 image sets of our study area, collected between 2022 and 2024.

While the SatlasPretrain ResNet model can easily detect road coverage, major bodies of water, general grass and shrubs and built-up areas, it does not capture the more complex and small scale agricultural land cover categories of rice paddies and agroforestry. To be clear, these categories are not included in the training effort. The omission is perhaps due to a combination of lack of interest in these conditions and the fact that Sentinel-2 imagery simply does not offer the spatial resolution required to detect small scale landscape conditions.

## VII. Responding to the evolving GeoAI landscape

The trend towards large pre-trained models will likely accelerate, and it will impact GeoAI and its downstream use and interpretation. The term foundation models [37] has been coined to denote any model that is trained on broad data and fine-tuned to a wide range of downstream tasks. Some large generative AI language models have evolved to become such foundation models and now dominate the landscape of chatbots, homogenizing the production of language in its wake. Just as large language models impact and alter the use of language, large geospatial models will likely impact how landscapes are viewed, classified, represented and cared for. The scale and intensity with which these future large geospatial models will be deployed, and likely become default references, will impact smaller organizations in the majority world differently than large and well-funded organizations with the means to readily fine-tune complex large models to their own needs.

Technologies developed for a world of abundance do not transfer directly to a world of scarce resources. The most powerful GeoAI algorithms need to be adapted in order to be of use in the majority world and for small organizations with limited means. Our experiences suggest that in this very fluid field, a mixed approach is warranted. On the one hand, Wisnu can rely on simple procedures performed on local computers to get basic insights from open source data providers. The skill sets to do so have been shared with Wisnu members in a series of tutorials and remote workshops. This sharing and knowledge transfer process serves not only as a basis for capacity building, but also gives the Wisnu foundation the agency to tune the task at hand to their owns needs, on their terms. Making use of the best possible geospatial algorithms, packaged into efficient pathways and tuned to the resource constraints of a given context, might be considered a specific variant of GeoAI, *small GeoAI*. As such, small GeoAI optimizes geospatial compute resources for specific stakeholders with particular needs. Additionally, it serves as a case study for computing within limits, a condition that is likely to receive more attention as energy use in high-end AI operations is scrutinized.

The inevitable arrival of homogenizing large models will require adaptation and new strategies. It is likely that small organizations will be at a disadvantage at first, as these models will likely not address their needs. However, as GeoAI research expands, the need for more differentiated datasets will become apparent where self-supervised systems fail to establish nuance and remain blind to deep local knowledge. At that point, organizations already familiar with GeoAI logics will be able to profit from the need for better training sets by perhaps selling their knowledge and the data that describes it to large AI organizations. This effort could be augmented by pooling needs and knowledge across NGOs invested in GeoAI, for example. Fostering a new generation of technology savvy contributors to GeoAI-supported land care is a key component, a requirement that is not always easy to meet. In Bali, for example, many young and talented individuals will opt to work in the tourism sector that offers better pay than to work for an NGO as an applied engineer.

Where resources are scarce, the appropriateness and effectiveness of end-to-end GeoAI pathways are more significant than the performance of individual GeoAI algorithms. New technology barriers occur where cloud-centric analysis systems and increasingly large and computationally expensive homogenizing foundation models come to constitute defaults, a condition that impacts not only NGOs but many smaller organizations. Over the long term, new alliances between compute-rich and compute-compromised regions and organizations should be established. Including more local knowledge from underrepresented regions of the world into curated data for training geospatial foundation models can contribute to making these models less blind to the richness of landscapes across the planet.

## Acknowledgments

This work was made possible in part by grants from Planet Labs and Google Research. Dr. Rajif Iryadi from the Indonesia Badan Riset dan Inovasi National offered invaluable support in parsing our satellite data. We furthermore acknowledge the contributions of the open source and QGIS communities.

## Supplementary materials

Code and data used in our experiments are available on GitHub:
https://github.com/realtechsupport/cocktail/tree/main/code/
https://github.com/realtechsupport/cocktail/tree/main/sandbox/working_model/